\documentstyle[preprint,aps,eqsecnum,axodraw]{revtex}

\newcommand{\beq}{\begin{equation}}
\newcommand{\eeq}{\end{equation}}
\newcommand{\snu}{\tilde \nu}

\newcommand{\msnuone}{m_{\snu_+}}
\newcommand{\msnutwo}{m_{\snu_-}}
\newcommand{\msnu}{m_{\snu}}

\newcommand{\dmsnutwo}{{\mbox{$\Delta m^2_{\snu}$}}}
\newcommand{\gsim}{\gtrsim}
\newcommand{\lsim}{\lesssim}

\newenvironment{Eqnarray}%
         {\arraycolsep 0.14em\begin{eqnarray}}{\end{eqnarray}}
\def\beqa{\begin{Eqnarray}}
\def\eeqa{\end{Eqnarray}}
\def\mz{m_Z}
\def\hh{H^0}
\def\hl{h^0}
\def\ha{A^0}
\def\mhh{m_{\hh}}
\def\mhl{m_{\hl}}
\def\mha{m_{\ha}}
\def\msnusnuab{(M^2_{\snu\snu^*})_{\alpha\beta}}
\def\msnusnumn{(M^2_{\snu\snu^*})_{mn}}

\def\l{\lambda}
\def\ls{\lambda^*}
\def\lp{\lambda^\prime}
\def\lps{\lambda^{\prime*}}
\def\ifmath#1{\relax\ifmmode #1\else $#1$\fi}
\def\eighth{\ifmath{{\textstyle{1 \over 8}}}}
\def\thsec{\ifmath{{\textstyle{1 \over 32}}}}
\def\quarter{\ifmath{{\textstyle{1 \over 4}}}}

\def\qslash{q\!\!\!\slash}
\def\pslash{p\!\!\!\slash}
\def\refs#1#2{refs.~\cite{#1} and \cite{#2}}
\def\Ref#1{ref.~\cite{#1}}

\def\eq#1{eq.~(\ref{#1})}
\def\Eq#1{Eq.~(\ref{#1})}
\def\eqs#1#2{eqs.~(\ref{#1}) and (\ref{#2})}
\def\eqss#1#2#3{eqs.~(\ref{#1}), (\ref{#2}) and (\ref{#3})}
\def\npb#1{{\sl Nucl.\ Phys.}\ {\bf B#1}}
\def\plb#1{{\sl Phys.\ Lett.}\ {\bf B#1}}
\def\prd#1{{\sl Phys.\ Rev.}\ {\bf D#1}}
\def\prl#1{{\sl Phys.\ Rev.\ Lett.} {\bf #1}}

\def\epjc#1{{\sl Eur.~Phys.~J.}\ {\bf C#1}}

\def\ifmath#1{\relax\ifmmode #1\else $#1$\fi}
\def\half{\ifmath{{\textstyle{1 \over 2}}}}
\def\sqhalf{\ifmath{{\textstyle{1 \over \sqrt{2}}}}}
\def\eighth{\ifmath{{\textstyle{1 \over 8}}}}
\def\cw{c_W}
\def\sw{s_W}

\begin{document}

\draft{\tighten

\preprint{
\vbox{
      \hbox{SLAC-PUB-7853}
      \hbox{SCIPP-98/31}
      \hbox{FERMILAB-PUB-98/345-T}
      \hbox{hep-ph/9810536}
      \hbox{October 1998}
    }}

\bigskip
\bigskip

\title{(S)neutrino properties in R-parity violating supersymmetry:\\
I. CP-conserving Phenomena}
\author{Yuval Grossman\,$^a$ and  Howard E. Haber\,$^{b,c}$}
\address{
$^a$Stanford Linear Accelerator Center, 
        Stanford University, Stanford, CA 94309 \\
  $^b$Santa Cruz Institute for Particle Physics, 
University of California, Santa Cruz, CA 95064 \\
$^c$Fermi National Accelerator Laboratory, P.O. Box 500, Batavia, IL 60510}

\maketitle

\begin{abstract}%
R-parity-violating supersymmetry (with a conserved baryon number $B$)
provides a framework for
particle physics with lepton number ($L$) violating interactions.  
We examine in detail the structure of the most 
general R-parity-violating ($B$-conserving) model of low-energy
supersymmetry.  We analyze
the mixing of Higgs bosons with sleptons and the mixing of charginos
and neutralinos with charged leptons and neutrinos, respectively.
Implications for neutrino and sneutrino masses and mixing and
CP-conserving sneutrino phenomena are considered.
$L$-violating low-energy supersymmetry can be probed at future 
colliders by studying the phenomenology of sneutrinos.
Sneutrino--antisneutrino mass splittings and lifetime differences can provide
new opportunities to probe lepton number violation at colliders.

\end{abstract}

}

\newpage

\renewcommand{\thefootnote}{\alph{footnote}}

\section{Introduction}
There is no fundamental principle 
that requires the theory of elementary particle interactions
to conserve lepton number.  In the Standard Model, lepton number
conservation is a fortuitous accident that arises because one cannot
write down renormalizable lepton-number-violating interactions that
only involve the fields of the Standard Model \cite{SteveW}.  In fact,
there are some experimental hints for non-zero neutrino masses \cite{pdg}
that suggest that lepton number is not an exact symmetry.

In low-energy supersymmetric extensions of the Standard Model, lepton
number conservation is not automatically respected by the most general
set of renormalizable interactions.  Nevertheless, 
experimental observations imply that lepton number
violating effects, if they exist, must be rather small.
If one wants to enforce lepton number
conservation in the tree-level supersymmetric theory, it is sufficient
to impose one extra discrete symmetry.  In the minimal
supersymmetric standard model (MSSM), a multiplicative symmetry called
R-parity is introduced, such that the R quantum number of
an MSSM field of spin $S$, baryon number $B$ and lepton number $L$ 
is given by $(-1)^{[3(B-L)+2S]}$. By introducing \hbox{$B\!\!-\!\!L$}
conservation modulo 2,  one eliminates all dimension-four lepton number 
and baryon number-violating interactions.  Majorana neutrino masses can
be generated in an R-parity-conserving extension of the MSSM 
involving new $\Delta L=2$ interactions through
the supersymmetric see-saw mechanism \cite{susyseesaw,GH}.  

In a recent paper \cite{GH} (for an independent study see \Ref{HKK}),
we studied the effect of such $\Delta L=2$ interaction on sneutrino phenomena.
In this case, the sneutrino ($\snu$) and antisneutrino ($\bar{\snu}$),
which are eigenstates of lepton number, are no longer mass eigenstates.  
The mass eigenstates are therefore superpositions of $\snu$ and $\bar{\snu}$,
and sneutrino mixing effects can lead to a phenomenology analogous to
that of $K$--$\overline K$ and $B$--$\overline B$ mixing. 
The mass splitting between the two sneutrino mass eigenstates is
related to the
magnitude of lepton number violation, which is typically characterized
by the size of neutrino masses.\footnote{In some cases the sneutrino mass 
splitting may be enhanced by a factor as large as $10^3$ compared 
to the neutrino mass \cite{GH,HMM}.}
As a result, the sneutrino mass splitting is expected generally to be
very small. Yet, it can be detected
in many cases, if one is able to observe the lepton number 
oscillation \cite{GH}.

Neutrino masses can
also be generated in R-parity-violating (RPV) models of low-energy
supersymmetry \cite{all,bgnn,enrico,Hem,Ferr}.  
However, all possible dimension-four RPV interactions
cannot be simultaneously present and unsuppressed; otherwise
the proton decay rate would be many orders of magnitude larger than
the present experimental bound.   One way to avoid proton decay is to
impose either $B$ or $L$ separately.  For example, if $B$ is conserved but $L$
is not, then the theory would
violate R-parity but preserve a ${\bf Z}_3$ baryon ``triality''.

In this paper we extend the analysis of \Ref{GH} and study 
sneutrino phenomena in models without R-parity (but with baryon triality). 
Such models exhibit
$\Delta L=1$ violating interactions at the level of renormalizable 
operators.  One can then generate $\Delta L=2$ violating interactions,
which are responsible for generating neutrino masses.
In general, one neutrino mass is generated at tree level via mixing with the 
neutralinos, and the remaining neutrino masses are generated at one-loop.

In Section II, we introduce the most general RPV model 
with a conserved baryon number and establish our notation. 
In Section III, we obtain the general form for the mass matrix in the
neutral fermion sector (which
governs the mixing of neutralinos and neutrinos) and in the
neutral scalar sector (which governs the mixing of neutral Higgs
bosons and sneutrinos).  From these results, we obtain the
tree-level masses of neutrinos and squared-mass splittings of the 
sneutrino--antisneutrino pairs.  In Section IV, we calculate the 
neutrino masses and sneutrino--antisneutrino squared-mass splittings
generated at one loop.  
The phenomenological implications of these results are addressed in
Section V along with our summary and conclusions.  
An explicit computation of the scalar potential of the
model is presented in Appendix A.  For completeness, we present 
in Appendix B the 
general form for the mass matrix in the charged fermion sector (which
governs the mixing of charginos and charged leptons) and in the
charged scalar sector (which governs the mixing of charged Higgs
bosons and charged sleptons).  The relevant Feynman rules for the RPV
model and the loop function needed for the one-loop computations of 
Section IV are given in Appendices C and D.

\section{R-parity violation formalism}
In R-parity-violating (RPV) low-energy supersymmetry, there is no conserved
quantum number that distinguishes the lepton supermultiplets $\hat
L_m$ and the down-type Higgs supermultiplet $\hat H_D$.  Here,
$m$ is a generation label that runs from 1 to $n_g=3$.  Each
supermultiplet transforms as a $Y=-1$ weak doublet under the electroweak
gauge group.  It is therefore convenient to denote the four
supermultiplets by one symbol $\hat L_\alpha$
($\alpha=0,1,\ldots,n_g$), with $\hat L_0\equiv \hat H_D$.  We consider the
most general low-energy supersymmetric model consisting of the
MSSM fields that conserves a ${\bf Z_3}$ baryon triality.  As remarked
in Section I, such a theory possesses RPV-interactions that violate
lepton number.

The Lagrangian of the theory is fixed by the superpotential and the
soft-supersymmetry-breaking terms (supersymmetry and gauge invariance
fix the remaining dimension-four terms).  The theory we consider
consists of the fields of the MSSM, {\it i.e} the fields of the
two-Higgs-doublet extension of the Standard Model plus their
superpartners.  The most general renormalizable superpotential
respecting baryon triality is given by:
\beq \label{rpvsuppot}
W=\epsilon_{ij} \left[
-\mu_\alpha \hat L_\alpha^i \hat H_U^j + 
\half\l_{\alpha\beta m}\hat L_\alpha^i \hat L_\beta^j \hat E_m +
\lp_{\alpha nm} \hat L_\alpha^i \hat Q_n^j  \hat D_m
-h_{nm}\hat H_U^i \hat Q^j_n \hat U_m
\right]\,,
\eeq
where $\hat H_U$ is the up-type Higgs supermultiplet, the
$\hat Q_n$ are doublet quark supermultipletss, $\hat U_m$ [$\hat D_m$] are 
singlet up-type [down-type] quark supermultiplets
and the $\hat E_m$ are the singlet charged lepton 
supermultiplets.\footnote{In our notation, $\epsilon_{12}=-\epsilon_{21}=1$.
The notation for the superfields (extended to allow $\alpha=0$
as discussed above) follows that of \Ref{habertasi}.
For example, $(\widetilde e_L^-)_m$ [$(\widetilde e_R^+)_m$] are the scalar
components of $\widehat L^2_m$ [$\widehat E_m$], {\it etc.}}  
Without loss of generality, the coefficients $\lambda_{\alpha\beta m}$ 
are taken to be antisymmetric under the interchange of the indices
$\alpha$ and $\beta$. Note that   
the $\mu$-term of the MSSM [which corresponds to $\mu_0$ in
\eq{rpvsuppot}]
is now extended to an $(n_g+1)$-component vector, $\mu_\alpha$
(while the latin indices $n$ and $m$ run from 1 to $n_g$). 
Then, the trilinear terms in the superpotential proportional to
$\lambda$ and $\lambda'$
contain lepton number violating generalizations of the down quark
and charged lepton Yukawa matrices.

Next, we consider the most general set of (renormalizable) 
soft-supersymmetry-breaking terms.  In addition to the usual 
soft-supersymmetry-breaking terms of the R-parity-conserving MSSM, one
must also add new $A$ and $B$ terms corresponding to the RPV
terms of the superpotential.  In addition, new RPV scalar squared-mass
terms also exist.  As above, we can streamline the notation by
extending the definitions of the coefficients of the R-parity-conserving
soft-supersymmetry-breaking terms to allow for an index of type
$\alpha$ which can run from 0 to $n_g$.  Explicitly,
\beqa \label{softsusy}
 V_{\rm soft}  &=&  (M^2_{\widetilde Q})_{mn}\,\widetilde Q^{i*}_m
        \widetilde Q^i_n 
   +  (M^2_{\widetilde U})_{mn}\,\widetilde U_m^*\widetilde U_n        
   +  (M^2_{\widetilde D})_{mn}\,\widetilde D_m^*\widetilde D_n
            \nonumber \\
 && +  (M^2_{\widetilde L})_{\alpha\beta}\,
          \widetilde L^{i*}_\alpha\widetilde L^i_\beta
      + (M^2_{\widetilde E})_{mn}\,\widetilde E_m^*\widetilde E_n
       + m^2_U|H_U|^2      
  -(\epsilon_{ij} b_\alpha\tilde L_\alpha^i H_U^j +{\rm h.c.}) \nonumber\\
 && +  \epsilon_{ij} \bigl[\half a_{\alpha\beta m} \widetilde L^i_\alpha
       \widetilde L^j_\beta \widetilde E_m + a'_{\alpha nm}
       \widetilde L^i_\alpha\widetilde Q^j_n\widetilde D_m
       - (a_U)_{nm} H^i_U
        \widetilde Q^j_n\widetilde U_m + {\rm h.c.}\bigr]\nonumber \\
 && +  \half \left[ M_3\, \widetilde g
   \,\widetilde g + M_2 \widetilde W^a\widetilde W^a
  + M_1 \widetilde B \widetilde B +{\rm h.c.}\right]\,.
\eeqa
Note that the
single $B$ term of the MSSM
is extended to an $(n_g+1)$-component vector, $b_\alpha$,
the single squared-mass term for the
down-type Higgs boson and the $n_g\times n_g$ lepton scalar squared-mass 
matrix are combined into an $(n_g+1)\times (n_g+1)$ matrix, 
and the matrix $A$-parameters
of the MSSM are extended in the obvious manner [analogous to the
Yukawa coupling matrices in \eq{rpvsuppot}].  In particular, 
$a_{\alpha\beta m}$ is antisymmetric under the interchange of
$\alpha$ and $\beta$.  It is sometimes 
convenient to follow the more conventional notation in the literature
and define the $A$ and $B$ parameters as follows:
\beqa \label{abterms}
&&a_{\alpha\beta m}\equiv \lambda_{\alpha\beta m} (A_E)_{\alpha\beta m}
\,,\qquad
(a_U)_{nm}\equiv h_{nm} (A_U)_{nm}\,,\nonumber \\
&&a'_{\alpha nm}\equiv \lambda'_{\alpha nm} (A_D)_{\alpha nm}\,,\qquad
b_\alpha\equiv \mu_\alpha B_\alpha\,,
\eeqa
where repeated indices are not summed over in the above equations.
Finally, the Majorana gaugino
masses, $M_i$, are unchanged from the MSSM.

The total scalar potential is given by:
\beq \label{scalarpot}
V_{\rm scalar}=V_F+V_D+V_{\rm soft}\,.
\eeq
In Appendix A, we present the complete expressions for $V_F$ (which is
derived from the superpotential [\eq{rpvsuppot}]) and $V_D$.  
It is convenient to write out the contribution of the neutral scalar
fields to the full scalar potential [\eq{scalarpot}]:
\beqa \label{Vn}
V_{\rm neutral} &=& 
\left(m_U^2+ |\mu|^2\right) {|h_U|}^2 + 
\left[({M^2_{\tilde L}})_{\alpha\beta} + \mu_\alpha\mu_\beta^*\right]
\snu_\alpha \snu_\beta^{*}
- \left(b_\alpha\snu_\alpha h_U +b_\alpha^*
\snu_\alpha^* h_U^*\right)  \\
&&\quad +\eighth(g^2+g^{\prime 2})
\left[|h_U|^2-|\snu_\alpha|^2 \right]^2\,,
\nonumber
\eeqa
where $h_U\equiv H_U^2$ is the neutral component of the up-type Higgs scalar
doublet and $\snu_\alpha\equiv \widetilde L^1_\alpha$.  
In \eq{Vn}, we have introduced the notation:
\beq \label{mudef}
|\mu|^2\equiv \sum_\alpha|\mu_\alpha|^2\,.
\eeq

In minimizing the full scalar potential, we assume that only neutral
scalar fields acquire vacuum expectation values: $\langle
h_U\rangle\equiv \sqhalf v_u$ and $\langle\snu_\alpha\rangle\equiv 
\sqhalf v_\alpha$.  From \eq{Vn}, the minimization conditions are:
\beqa 
(m_U^2+|\mu|^2)v_u^* &=& b_\alpha v_\alpha-\eighth(g^2+g^{\prime 2})
(|v_u|^2-|v_d|^2)v_u^*
\,,\label{mincondsa} \\
\left[({M^2_{\tilde L}})_{\alpha\beta} + \mu_\alpha\mu_\beta^*\right]
v_\beta^* &=& b_\alpha
v_u+\eighth(g^2+g^{\prime 2})(|v_u|^2-|v_d|^2)v_\alpha^*\,,\label{mincondsb} 
\eeqa
where   
\beq\label{vdef}
|v_d|^2\equiv \sum_\alpha |v_\alpha|^2\,.
\eeq
The normalization of the vacuum expectation values has been chosen
such that
\beq \label{vevdef}
v\equiv (|v_u|^2+|v_d|^2)^{1/2}={2m_W\over g}=246~{\rm GeV}\,.
\eeq

Up to this point, there is no preferred direction in the generalized
generation space spanned by the $\hat L_\alpha$.  It is convenient to
choose a particular ``interaction'' basis such that 
$v_m=0$ ($m=1,\ldots,n_g$), in which case $v_0=v_d$.
In this basis, we denote
$\hat L_0\equiv\hat H_D$.  The down-type quark and lepton mass 
matrices in this basis arise from the Yukawa couplings to $H_D$; 
namely,\footnote{As shown in Appendix B, $(m_\ell)_{nm}$ is not precisely
the charged lepton mass matrix, as a result of a small admixture of
the charged higgsino eigenstate due to RPV interactions.}
\beq \label{ellmassdef}
(m_d)_{nm} = \sqhalf v_d \lp_{0nm}\,, \qquad
(m_\ell)_{nm} = \sqhalf v_d \l_{0nm}\,,
\eeq
while the up-type quark mass matrices arise as in the MSSM:
\beq
(m_u)_{nm} = \sqhalf v_u h_{nm}\,. 
\eeq
In the literature, one often finds other basis choices.  
The most common is one where $\mu_0=\mu$ and $\mu_m=0$ ($m=1,\ldots,n_g$).  
Of course, the results for physical observables (which involve mass
eigenstates) are independent of the basis choice.\footnote{For a general
discussion of basis indpendent parameterizations of R-parity
violation, see \refs{sacha}{Ferr}.}  
In the calculations presented in this paper, 
when we need to fix a basis, we find the choice of $v_m=0$ to be 
the most convenient.

\section{Neutrinos and sneutrinos at tree level}
We begin by recalling the calculation of the tree-level neutrino mass 
that arises due to the R-parity violation. We then evaluate the
corresponding sneutrino mass splitting.  In all the subsequent analysis
presented in this paper, we shall assume for simplicity that the
parameters $(M^2_{\widetilde L})_{\alpha\beta}$, $\mu_\alpha$,
$b_\alpha$, the gaugino mass parameters $M_i$, and $v_\alpha$ are
real.  In particular, the ratio of vacuum expectation values,
\beq
\tan\beta\equiv {v_u\over v_d}
\eeq
can be chosen to be positive by convention [with $v_d$ defined by the
positive square root of \eq{vdef}].
That is, we neglect new supersymmetric sources of CP-violation
that can contribute to neutrino and sneutrino phenomena. 
We shall address the latter possibility in a subsequent 
paper \cite{ghprep}.

\subsection{Neutrino mass}

The neutrino can become massive due to mixing with the neutralinos \cite{all}.
This is determined by the $(n_g+4)\times (n_g+4)$ mass matrix 
in a basis spanned by the two
neutral gauginos $\widetilde B$ and $\widetilde W^3$, the higgsinos
$\widetilde h_U$ and $\widetilde h_D\equiv\nu_0$,
and $n_g$ generations of neutrinos, $\nu_m$.  The tree-level 
fermion mass matrix, with rows and columns corresponding to
$\{\widetilde B,\widetilde W^3,\widetilde h_U, 
\nu_\beta~(\beta=0,1,\ldots,n_g)\}$ is given by \cite{bgnn,enrico}:
\beq
M^{\rm (n)}=\pmatrix{
M_1&0&m_Z\sw v_u/v&-m_Z\sw v_\beta/v\cr
0&M_2&-m_Z\cw v_u/v&m_Z\cw v_\beta/v\cr
m_Z\sw v_u/v&-m_Z\cw v_u/v&0&\mu_\beta\cr
-m_Z\sw v_\alpha/v&m_Z\cw v_\alpha/v&\mu_\alpha&0_{\alpha\beta}\cr}\,,
\eeq
where $\cw\equiv\cos\theta_W$, $\sw\equiv\sin\theta_W$, $v$ is defined 
in \eq{vevdef}, and $0_{\alpha\beta}$ is the $(n_g+1)\times (n_g+1)$
zero matrix.  In a basis-independent analysis, 
it is convenient to introduce:
\beq\label{xidef}
\cos\xi\equiv {\sum_\alpha v_\alpha\mu_\alpha\over v_d\mu},
\eeq
where $\mu$ is defined in \eq{mudef}.  Note that $\xi$
measures the alignment of $v_\alpha$ and $\mu_\alpha$.
It is easy to check that $M^{(n)}$ possesses $n_g-1$ zero eigenvalues.
We shall identify the corresponding states with $n_g-1$
physical neutrinos of the Standard Model \cite{bgnn}, while
one neutrino acquires mass through mixing.
We can evaluate this mass by computing the product of the five
non-zero eigenvalues of $M^{(n)}$ [denoted below by
$\det' M^{\rm (n)}$]~\footnote{To compute this
quantity, calculate the characteristic polynomial, $\det(\lambda I-M^{(n)})$
and examine the first non-zero coefficient of $\lambda^n$
($n=0,1,\ldots$).  In the present case, $\det' M^{\rm (n)}$ is given by
the coefficient of $\lambda^{n_g-1}$.}
\beq
\mbox{$\det'$} M^{\rm (n)} = 
m_Z^2 \mu^2 M_{\tilde \gamma}\cos^2\beta \sin^2\xi\,,
\eeq
where $M_{\tilde \gamma}\equiv \cos^2\theta_W M_1 + \sin^2\theta_W M_2$.
We compare this result with the product of the four neutralino masses
of the R-parity-conserving MSSM (obtained by computing the determinant
of the upper $4\times 4$ block of $M^{(n)}$ with $\mu_0$, $v_0$ 
replaced by $\mu$, $v_d$ respectively)
\beq
\det M^{\rm (n)}_0 = 
\mu\left(m_Z^2 M_{\tilde \gamma}\sin 2\beta-M_1 M_2 \mu\right).
\eeq
To first order in the neutrino mass, the neutralino masses are unchanged
by the R-parity violating terms, and we end up with \cite{enrico}
\beq
m_\nu = {\det' M^{\rm (n)} \over \det M^{\rm (n)}_0} =
{m_Z^2 \mu M_{\tilde \gamma}\cos^2\beta \sin^2\xi
\over m_Z^2 M_{\tilde \gamma}\sin 2\beta-M_1 M_2 \mu}\,.
\eeq
Thus, $m_\nu \sim m_Z \cos^2\beta \sin^2\xi$, assuming that all the relevant 
masses are at the electroweak scale.

Note that a necessary and sufficient condition for $m_\nu\neq 0$ 
(at tree-level) is $\sin\xi\neq 0$, which implies that 
$\mu_\alpha$ and $v_\alpha$ are not aligned.  
This is generic in RPV models.
In particular, the alignment of $\mu_\alpha$ and $v_\alpha$ is not
renormalization group invariant\cite{enrico,Hem}.
Thus, exact alignment at the low-energy scale 
can only be implemented by the fine-tuning of the model parameters.

\subsection{Sneutrino mass splitting}
In RPV low-energy supersymmetry, the sneutrinos mix with the Higgs
bosons.  Under the assumption of CP-conservation, we may separately
consider the CP-even and CP-odd scalar sectors.  For simplicity,
consider first the case of one sneutrino generation. 
If R-parity is conserved, the CP-even scalar sector consists of two
Higgs scalars ($\hl$ and $\hh$, with $\mhl<\mhh$) and 
$\snu_+$, while the CP-odd scalar sector consists of 
the Higgs scalar, $\ha$, the Goldstone
boson (which is absorbed by the $Z$), 
and one sneutrino, $\snu_-$.  Moreover, the $\snu_\pm$ are mass
degenerate, so that the standard practice is to define eigenstates of
lepton number: $\snu\equiv (\snu_+ + i\snu_-)/\sqrt{2}$ and 
$\bar{\snu}\equiv\snu^*$.  When R-parity is violated, the sneutrinos
in each CP-sector mix with the corresponding Higgs scalars, and the
mass degeneracy of $\snu_+$ and $\snu_-$ is broken.
We expect the RPV-interactions to be small; thus, we can evaluate the 
concomitant sneutrino mass splitting in perturbation theory.
For $n_g>1$ generations of sneutrinos,
one can consider non-trivial flavor mixing among sneutrinos 
(or antisneutrinos) in addition to $n_g$ sneutrino--antisneutrino mass
splittings.  

The CP-even and CP-odd scalar squared-mass matrices 
are most easily derived as follows.
Insert $h_U=\sqhalf (v_u+ia_u)$
and $\snu_\alpha=\sqhalf (v_\alpha+ia_\alpha)$ into \eq{Vn} and
call the resulting expression $V_{\rm even}+V_{\rm odd}$.
The CP-even squared-mass matrix is obtained from $V_{\rm even}$, 
which is identified by replacing the scalar fields 
in \eq{Vn} by their corresponding real
vacuum expectation values (or equivalently by setting $a_u=a_\alpha=0$
in $V_{\rm even}+V_{\rm odd}$).  Then, 
\beqa
&&V_{\rm even}= \half m_{uu}^2 v_u^2 + \half m_{\alpha\beta}^2
v_\alpha v_\beta - b_\alpha v_u v_\alpha
+\thsec(g^2+g^{\prime 2})\left(v_u^2-v_d^2\right)^2\,,\label{veven}\\
&&V_{\rm odd}= \half m_{uu}^2 a_u^2 + \half m_{\alpha\beta}^2
a_\alpha a_\beta + b_\alpha a_u a_\alpha
+\thsec(g^2+g^{\prime 2})\left[(a_u^2-a_d^2)^2+2(a_u^2-a_d^2)(v_u^2-v_d^2)
\right],\nonumber \\ \label{vodd}
\eeqa
where $m_{uu}^2\equiv \left(m_U^2+\mu^2\right)$ and
$m_{\alpha\beta}^2\equiv ({M^2_{\tilde L}})_{\alpha\beta}+ 
\mu_\alpha\mu_\beta$.  
The minimization conditions 
$d V_{\rm even}/dv_p=0$ ($p=u,\alpha$) yield
\eqs{mincondsa}{mincondsb}, with all parameters assumed to be real.  
In particular, it is convenient to rewrite \eq{mincondsb}.
First, we introduce the 
generalized $(n_g+1)\times (n_g+1)$ sneutrino squared-mass matrix:
\beq \label{treelevelsnumasses}
\msnusnuab\equiv
({M^2_{\tilde L}})_{\alpha\beta}+ \mu_\alpha\mu_\beta-\eighth 
(g^2+g^{\prime 2})(v_u^2-v_d^2)\delta_{\alpha\beta}\,.
\eeq
Then, \eq{mincondsb} assumes a very simple form:
\beq \label{veveq}
\msnusnuab v_\beta=v_u b_\alpha\,.
\eeq

{}From this equation, we can derive the necessary and sufficient
condition for $\sin\xi=0$ (corresponding to the
alignment of $\mu_\alpha$ and $v_\alpha$). 
If there exist some number $c$ such that
\beq \label{align}
\msnusnuab\mu_\beta=c\,b_\alpha\,,
\eeq
then it follows that $\mu_\alpha$ and $v_\alpha$ are 
aligned.\footnote{It is
interesting to compare this result with the one obtained in
\Ref{bgnn}, where it was shown that $\mu_\alpha$ and
$v_\alpha$ are aligned if
two conditions hold: (i) $b_\alpha \propto\mu_\alpha$ and (ii)
$\mu_\alpha$ is an eigenvector of $(M^2_{\widetilde L})_{\alpha\beta}$.
{}From \eq{align}, we see that these two conditions are sufficient for
alignment [since conditions (i) and (ii) imply the existence of
a constant $c$ in \eq{align}], but are not the most general.}
To prove that \eq{align} implies the alignment of $\mu_\alpha$ and
$v_\alpha$, simply insert \eq{align} into \eq{veveq} [thereby
eliminating $b_\alpha$], and note that $\msnusnuab$ must
be non-singular [otherwise \eq{veveq} would not yield
a unique non-trivial solution for $v_\alpha$].

Naively, one might think that if $\mu_\alpha$ and $v_\alpha$ are aligned,
so that all tree-level neutrino masses
vanish, then one would also find degenerate sneutrino--antisneutrino pairs
at tree-level.  This is not generally true.  Instead, the absence
of degenerate sneutrino--antisneutrino pairs is controlled by the
alignment of $b_\alpha$ and $v_\alpha$.  To see how this works, 
note that \eq{veveq} implies that $b_\alpha$ and $v_\alpha$ are
aligned if $v_\beta$ is an eigenvector of
$\msnusnuab$.  In
this case, one can rotate to a basis in which $v_m=b_m=0$ (where
$m=1,\ldots, n_g$).  In this basis the matrix elements
$(M^2_{\snu\snu^*})_{0m}=(M^2_{\snu\snu^*})_{m0}=0$, 
which implies that there is no mixing between
Higgs bosons and sneutrinos.  Thus, although some RPV effects still
remain in the theory, the CP-even and CP-odd sneutrino mass matrices 
are identical.  Consequently, the conditions for the absence of
tree-level neutrino masses (alignment of $\mu_\alpha$ and $v_\alpha$)
and the absence of sneutrino--antisneutrino mass
splitting at tree-level (alignment of $b_\alpha$ and $v_\alpha$)
are different.  

To compute the tree-level sneutrino--antisneutrino mass splittings,
we must calculate the CP-even and CP-odd scalar spectrum.  The CP-even
scalar squared-mass matrix is given by
\beq
(M^2_{\rm even})_{pq} = {d^2 V_{\rm even} \over dv_p dv_q}\,.
\eeq 
After using the minimization conditions of the potential, 
we obtain the following result for the CP-even
squared-mass matrix
\beq \label{meven2}
M_{\rm even}^2= \pmatrix{
\quarter(g^2+g^{\prime 2})v_u^2+b_\rho v_\rho/v_u&
-\quarter(g^2+g^{\prime 2})v_u v_\beta -b_\beta\cr
-\quarter(g^2+g^{\prime 2})v_u v_\alpha -b_\alpha &
\quarter(g^2+g^{\prime2})v_\alpha v_\beta +\msnusnuab\cr}\,,
\eeq
where $\msnusnuab$ is constrained according to \eq{veveq}.
The CP-odd scalar squared-mass matrix is determined from 
\beq
(M^2_{\rm odd})_{pq} = \left.{d^2 V_{\rm odd} \over da_p da_q}
\right|_{a_p=0}\,,
\eeq 
where $V_{\rm odd}$ is given by \eq{vodd}.  The resulting CP-odd
squared-mass matrix is then
\beq \label{modd2}
M_{\rm odd}^2= \pmatrix{
b_\rho v_\rho/v_u & b_\beta \cr
b_\alpha & \msnusnuab \cr}\,.
\eeq
Note that the vector $(-v_u,v_\beta)$ is an eigenvector of $M_{\rm
odd}^2$ with zero eigenvalue; this is the Goldstone boson that is
absorbed by the $Z$.  One can check that the following 
tree-level sum rule holds:
\beq \label{evenoddsumrule}
{\rm Tr}~M^2_{\rm even}= m_Z^2+{\rm Tr}~M^2_{\rm odd}\,.
\eeq
This result is a generalization of the well known tree-level sum rule
for the CP-even Higgs masses of the MSSM [see \eq{higgsrela}].
\Eq{evenoddsumrule} is more general in that it also includes
contributions from the sneutrinos which mix with the neutral Higgs
bosons in the presence of RPV interactions. 

To complete the computation of the sneutrino--antisneutrino 
mass splitting, one must evaluate the non-zero eigenvalues of
$M_{\rm even}^2$ and $M_{\rm odd}^2$, and identify which ones correspond
to the sneutrino eigenstates.  To do this, one must first
identify the small parameters characteristic of the RPV interactions.
We find that a judicious choice of basis significantly simplifies
the analysis.  Following the discussion at the end of Section II, we
choose a basis such that $v_m=0$ (which implies that $v_d=v_0$).

To illustrate our method, we exhibit the calculation in the case of
$n_g=1$ generation. In the basis where $v_1=0$, \eq{veveq}
implies that $(M^2_{\snu\snu^*})_{\alpha 0}=b_\alpha\tan\beta$
($\alpha=0,1$).  Then the squared-mass matrices
\eqs{meven2}{modd2} reduce to:
\beq \label{meven2sp}
M_{\rm even}^2= \pmatrix{
b_0\cot\beta+\quarter(g^2+g^{\prime 2})v_u^2 & 
-b_0-\quarter(g^2+g^{\prime 2}) v_u v_d & -b_1 \cr
-b_0-\quarter(g^2+g^{\prime 2}) v_u v_d &
b_0\tan\beta+\quarter(g^2+g^{\prime 2})v_d^2 & b_1\tan\beta \cr
-b_1 &  b_1\tan\beta & m_{\snu\snu^*}^2 \cr}\,,
\eeq
and
\beq \label{modd2sp}
M_{\rm odd}^2= \pmatrix{
b_0\cot\beta & b_0 & b_1 \cr
b_0 & b_0\tan\beta & b_1\tan\beta \cr
b_1 &  b_1\tan\beta & m_{\snu\snu^*}^2 \cr}\,,
\eeq
where 
\beq\label{moredefs}
m_{\snu\snu^*}^2\equiv (M^2_{\snu\snu^*})_{11}=
(M^2_{\tilde L})_{11}+\mu_1^2-\eighth (g^2+g^{\prime 2})(v_u^2-v_d^2)\,.
\eeq
In the R-parity-conserving limit ($b_1=\mu_1=0$), one obtains the
usual MSSM tree-level masses for the Higgs bosons and the sneutrinos.

In both squared-mass matrices [\eqs{meven2sp}{modd2sp}], 
$b_1 \ll m_Z^2$ is a small parameter that can be treated
perturbatively.  We may then compute the sneutrino mass splitting
due to the small mixing with the Higgs bosons.  Using second order matrix 
perturbation theory to compute the eigenvalues, we find:
\beqa \label{snuonesnutwo}
&&\msnuone^2=m_{\snu\snu^*}^2+{b_1^2\over \cos^2\beta}\left[
{\sin^2(\beta-\alpha) \over (m_{\snu\snu^*}^2-\mhh^2)} + 
{\cos^2(\beta-\alpha) \over
(m_{\snu\snu^*}^2-\mhl^2)}\right]\,, \nonumber \\
&&\msnutwo^2=m_{\snu\snu^*}^2+
{b_1^2 \over (m_{\snu\snu^*}^2-\mha^2)\cos^2\beta}\,. 
\eeqa
Above, we employ the standard notation for the MSSM Higgs sector
observables \cite{hunter}.  Note that at leading order in
$b_1^2$, it suffices to use the values for the Higgs parameters in
the R-parity-conserving limit.  In particular, the (tree-level) Higgs
masses satisfy:
\beqa 
&&\mhl^2+\mhh^2=m_Z^2+\mha^2\,, \label{higgsrela}\\
&&\mhl^2 \mhh^2 = m_Z^2 \mha^2 \cos^2 2\beta \label{higgsrelb}\,,
\eeqa
while the (tree-level) CP-even Higgs mixing angle satisfies: 
\beq
\cos^2(\beta-\alpha)={\mhl^2(\mz^2-\mhl^2)\over
\mha^2(\mhh^2-\mhl^2)}\,. 
\eeq
After some algebra, we end up with the following expression at leading
order in $b_1^2$ for the sneutrino squared-mass splitting,
$\dmsnutwo\equiv \msnuone^2-\msnutwo^2$:
\beq \label{dms}
\dmsnutwo = {4 \, b_1^2\, m_Z^2 \, m_{\snu\snu^*}^2 \, \sin^2\beta \over
(m_{\snu\snu^*}^2-m_H^2) (m_{\snu\snu^*}^2-m_h^2) (m_{\snu\snu^*}^2-m_A^2)}\,.
\eeq

We now extend the above results to more than one generation of
sneutrinos.  In a basis where $v_m=0$ ($m=1,\ldots, n_g$),
the resulting CP-even and CP-odd squared mass matrices
are obtained from \eqs{meven2sp}{modd2sp} by replacing $b_1$ with
the $n_g$-dimensional vector $b_m$ and
$m^2_{\snu\snu^*}$ by the $n_g \times n_g$ matrix,
$\msnusnumn$.
In general, $\msnusnumn$ need not be flavor diagonal.  In this
case, the theory would predict sneutrino flavor mixing in addition to
the sneutrino--antisneutrino mixing exhibited above.  The relative
strength of these effects depends on the relative size of the RPV and
flavor-violating parameters of the model.  To analyze the resulting
sneutrino spectrum, we choose a basis in which $\msnusnumn$ is
diagonal:
\beq \label{diagsnumass}
\msnusnumn=(m^2_{\snu\snu^*})_m\delta_{mn}\,.
\eeq
In this basis $b_m$ is also suitably redefined.
(We will continue to use the same symbols for these quantities in the
new basis.)  The CP-even and CP-odd sneutrino mass 
eigenstates will be denoted by $(\snu_{+})_m$ and
$(\snu_{-})_m$ respectively.\footnote{The index $m$ 
labels sneutrino generation, although one
should keep in mind that in the presence of flavor violation, the
sneutrino mass basis is not aligned with the corresponding mass bases 
relevant for the charged sleptons, charged leptons, or neutrinos.}
It is a simple matter to extend the perturbative analysis
of the scalar squared-mass matrices if the 
$(m^2_{\snu\snu^*})_m$ are non-degenerate.  We then find that 
$(\dmsnutwo)_m \equiv  (\msnuone^2)_m- (\msnutwo^2)_m$ is given by
\eq{dms}, with the replacement of $b_1$ and $m^2_{\snu\snu^*}$ by 
$b_m$ and $(m^2_{\snu\snu^*})_m$, respectively.  That is,
while in general only one neutrino is massive, 
all the sneutrino--antisneutrino pairs are generically 
split in mass.\footnote{This is a very general tree-level result.  Consider
models with $n_g$ generations of left-handed neutrinos in which
some of the neutrino mass eigenstates remain massless.
One finds that generically, {\it all} $n_g$
sneutrino--antisneutrino pairs are split in mass. 
For example, in the three generation see-saw model with
one right handed neutrino, only one neutrino is massive, while all 
three sneutrino--antisneutrino pairs are non-degenerate.
(At the one-loop level, the non-degeneracy of the
sneutrino-antisneutrino pairs will generate small masses for neutrinos
that were massless at tree level \cite{davking}.)} 
If we are prepared to allow for special choices of the parameters
$\mu_\alpha$ and $b_\alpha$, then these results are modified.  The one
massive neutrino becomes massless if $\mu_m=0$ for {\it all} $m$
(in the basis where $v_m=0$).  In contrast, the number of
sneutrino--antisneutrino pairs that remain degenerate in mass is equal
to the number of the $b_m$ that are zero. (Of course,
all these tree-level results 
are modified by one loop radiative corrections as discussed in Section IV.)

If some of the $(m^2_{\snu\snu^*})_m$ are degenerate, the analysis
becomes significantly more complicated.  We will not provide the
corresponding analytic expressions (although they can be obtained
using degenerate second order perturbation theory).  However, one can
show that for two or more generations
if $n_{\rm deg}$ of the $(m^2_{\snu\snu^*})_m$ are equal (by definition,
$n_{\rm deg}\geq 2$), and if $b_m\neq 0$ for all $m$
then only $n_g-n_{\rm deg}+2$ of the CP-even/CP-odd sneutrino pairs are 
split in mass.  The remaining $n_{\rm deg}-2$ sneutrino pairs are exactly
mass-degenerate at tree-level.  Additional cases can be
considered if some of the $b_m$ vanish.  

\section{One-loop effects}
In Section III, we showed that in the three generation model
for a generic choice of RPV-parameters,
mass for one neutrino flavor is generated at tree-level due to mixing
with the neutralinos, while mass splittings of three generations of
sneutrino--antisneutrino pairs at tree level are a consequence of 
mixing with the Higgs bosons.  Special choices of the RPV parameters
can leave all neutrinos massless at tree-level and/or less than 
three sneutrino--antisneutrino pairs with non-degenerate tree-level masses.

Masses for the remaining massless neutrinos and mass splittings for
the remaining degenerate sneutrino--antisneutrino pairs
will be generated by one loop effects.  Moreover, in some cases, the
radiative corrections to the tree-level generated masses and
mass splittings can be
significant (and may actually dominate the corresponding tree-level results).
As a concrete example, consider a model in which RPV interactions are
introduced only through the superpotential $\lambda$ and $\lambda'$
couplings [\eq{rpvsuppot}].  
In this case, $\mu_\alpha$, $b_\alpha$ and $v_\alpha$ are all trivially
aligned and no tree-level neutrino masses nor sneutrino mass splittings
are generated.  In a realistic model, soft-supersymmetry-breaking RPV-terms
will be generated radiatively in such models, thereby introducing
a small non-alignment among $\mu_\alpha$, $v_\alpha$ and $b_\alpha$.
However, the resulting tree-level
neutrino masses and sneutrino--antisneutrino mass splittings will be
radiatively suppressed, in which case the tree-level and one loop radiatively
generated masses and mass splittings considered in this section would
be of the same order of magnitude.

In this section, we compute the one loop generated neutrino
mass and sneutrino--antisneutrino mass splitting generated by the
RPV interactions.  However, there is another effect that arises at
one loop from R-parity conserving effects.  Once a
sneutrino--antisneutrino squared-mass splitting is established, its
presence will contribute radiatively to neutrino masses through a
one loop diagram involving sneutrinos and neutralinos (with R-parity
conserving couplings).  Similarly, a non-zero neutrino mass will
generate a one loop sneutrino--antisneutrino mass splitting.
In \Ref{GH}, we considered these effects explicitly.  The conclusion
of this work was that 
\beq \label{rnulimits}
10^{-3}\lsim {\Delta\msnu\over m_\nu}\lsim 10^3\,.
\eeq
This result is applicable in all models in which there is no
unnatural cancellation
between the tree-level and one loop contribution to the neutrino mass
or to the sneutrino--antisneutrino mass splitting.

\subsection{One-loop Neutrino mass}
At one loop, contributions to the neutrino mass are generated 
from diagrams involving charged lepton-slepton loop
(shown in Fig.~1) and an analogous down-type
quark-squark loop \cite{all}.
We first consider the contribution of the charged lepton-slepton 
loop.  We shall work in a specific basis, in which $v_m=0$ ({\it i.e.},
$v_0 \equiv v_d$) 
and the charged lepton mass matrix is diagonal.  In this basis, the
distinction between charged sleptons and Higgs bosons is meaningful.
Nevertheless, in a complete calculation, we should keep track of
charged slepton--Higgs boson mixing and the charged lepton--chargino
mixing which determine the actual mass eigenstates that appear in the
loop.  For completeness, we write out in Appendix B the relevant
mass matrices of the charged fermion and scalar sectors. 
In order to simplify the computation, we shall simply ignore all
flavor mixing (this includes mixing between charged Higgs
bosons and sleptons).  However,
we allow for mixing between the L-type and R-type charged sleptons 
separately in each generation, since this is necessary in order to
obtain a non-vanishing effect.  

It therefore suffices to consider the structure
of a $2\times 2$ (LR) block of the charged slepton squared-mass matrix
corresponding to one generation. 
The corresponding charged slepton mass eigenstates are given by:
\beq
\widetilde\ell_i=V_{i1}\widetilde \ell_L+V_{i2}\widetilde \ell_R\,,
\qquad i=1,2\,,
\eeq
where
\beq \label{slmix}
V=\pmatrix{\cos\phi_\ell & \sin\phi_\ell \cr
-\sin\phi_\ell & \cos\phi_\ell \cr}\,.
\eeq
The mixing angle $\phi_\ell$ can be found by diagonalizing the
charged slepton squared-mass matrix
\beq \label{sllr}
M^2_{\rm slepton}=\pmatrix{ L^2+m_\ell^2 & A m_\ell \cr
A m_\ell & R^2+m_\ell^2 \cr},
\eeq
where $L^2\equiv (M^2_{\widetilde L})_{\ell\ell} + 
(T_3 - e \sin^2\theta_W) m_Z^2 \cos 2 \beta$,
$R^2\equiv (M^2_{\widetilde E})_{\ell\ell} + 
(e \sin^2\theta_W) m_Z^2 \cos 2 \beta$, 
with $T_3=-1/2$ and $e=-1$ for the down-type charged sleptons,
and $A\equiv (A_E)_{0\ell\ell}-\mu_0\tan\beta$.  
In terms of these parameters, the mixing angle is given by
\beq \label{sinphi}
\sin2\phi_\ell = {2 A m_\ell \over \sqrt {(L^2-R^2)^2+4 A^2 m_\ell^2}}.
\eeq

The two-point amplitude corresponding to Fig.~1 can be computed using
the Feynman rules given in Appendix C.  The result is
given by
\beq
i{\cal M}_{qm}=\sum_{\ell,p}\sum_{i=1,2}
\int {d^4 q \over (2 \pi)^4} (-i \lambda_{q\ell p}) C^{-1} P_L 
V_{i2} {i (\qslash + m_\ell) \over q^2-m_\ell^2} (i \lambda_{mp\ell})
P_L V_{i1} {i \over (q-p)^2-M_{p_i}^2}\,,
\eeq
where $m_\ell$ is the lepton mass, $M_{p_i}$ are the sleptons masses
and the $V_{ij}$ are the slepton mixing matrix elements [\eq{slmix}].
The charge conjugation matrix $C$ appears according to the Feynman rules
given in Appendix D of \Ref{haberkane}.
The integral above can be expressed in terms of the well known one 
loop-integral $B_0$ (defined in Appendix D).  The corresponding 
contributions to the one loop neutrino mass matrix is obtained via: 
$(m_\nu)_{qm}=-{\cal M}_{qm}({p^2=0})$.  The end result is 
\beqa \label{oneloopnu}
(m_\nu)_{qm}^{(\ell)}& =& {1 \over 32\pi^2}\sum_{\ell,p}
\lambda_{q\ell p} \lambda_{mp\ell} m_\ell\sin 2\phi_\ell 
\left[ B_0(0,m_n^2,M_{p_1}^2)-B_0(0,m_n^2,M_{p_2}^2) \right]\nonumber \\
&\simeq & {1 \over 32\pi^2}\sum_{\ell,p}
\lambda_{q\ell p} \lambda_{mp\ell} m_\ell \sin 2\phi_\ell
\ln\left({M_{p_1}^2 \over M_{p_2}^2}\right)\,,
\eeqa
where the superscript $(\ell)$ indicates the contribution of Fig.~1.
As expected, the divergences cancel and the final result is finite.
In the last step, we simplified the resulting expression under the
assumption that $m_\ell\ll M_{p_1}$, $M_{p_2}$.

\begin{center}
\begin{picture}(200,120)(0,0)
\ArrowArcn(100,60)(40,180,0)
\DashArrowArcn(100,60)(40,0,180)5
\ArrowLine(10,60)(60,60)
\ArrowLine(190,60)(140,60)
\Text(30,50)[]{$\nu_m(p)$}
\Text(160,50)[]{$\nu_q(p)$}
\Text(95,115)[]{$\ell^-_n(q)$}
\Text(95,9)[]{$\widetilde \ell^-_p(q-p)$}
\end{picture}
\end{center}
\centerline{Fig. 1. One-loop contribution to the neutrino mass.}
\vspace*{3mm}

The  quark-squark loop contribution to the one loop neutrino mass may
be similarly computed.  Employing the same approximations as described
above, the final result can be immediately obtained from 
\eq{oneloopnu} with the following adjustments: (i) multiply the result
by a color factor of $N_c=3$; (ii) replace the Yukawa couplings
$\lambda$ with $\lambda'$ and the lepton mass $m_\ell$ by the
corresponding down-type quark mass $m_d$; (iii) replace the slepton
mixing angle $\phi_\ell$ by the corresponding down-type squark mixing
angle $\phi_d$.  Note that $\phi_d$ is computed using
\eqs{slmix}{sinphi}, after replacing $m_\ell$, $e=-1$,
$M^2_{\widetilde L}$, $M^2_{\widetilde E}$ and
$(A_E)_{0\ell\ell}$ with $m_d$, $e=-1/3$, $M^2_{\widetilde Q}$,
$M^2_{\widetilde D}$, and $(A_D)_{0dd}$ respectively.  Here and
below, $d$ [$r$] labels the generations of down-type quarks [squarks].
Then, 
\beq \label{oneloopnu2}
(m_\nu)_{qm}^{(d)} \simeq  {3 \over 32\pi^2}\sum_{d,r}
\lambda'_{qdr} \lambda'_{mrd} m_d \sin 2\phi_d
\ln\left({M_{r_1}^2 \over M_{r_2}^2}\right)\,.
\eeq

The final result for the neutrino mass matrix is the sum of
\eqs{oneloopnu}{oneloopnu2}.  Clearly, for generic choices of the
$\lambda$ and $\lambda'$ couplings, all neutrinos (including those 
neutrinos that were massless at tree-level)
gain a one loop generated mass.

\subsection{One-loop sneutrino-antisneutrino mass splitting}


We next consider the computation of the
one-loop contributions to the sneutrino masses 
under some simplifying assumptions (which are sufficient to illustrate
the general form of these corrections).
Since the total R-parity conserving contribution to the sneutrino and
antisneutrino mass is equal and large (of order the supersymmetry
breaking mass), it is sufficient to evaluate the one loop corrections
to the $\Delta L=2$ sneutrino squared-masses.
Flavor non-diagonal contributions are 
significant only if sneutrinos of different flavors are mass-degenerate.
The one loop generated mass splitting is relevant only when the tree 
level contributions vanish or are highly suppressed. 
In the simplest case, for one generation of sneutrinos and without
tree-level sneutrino--antisneutrino splitting, we get
\beq
(\Delta m^2_{\snu})_n = 2\left|{\cal M}_{nn}(p^2=m^2_{\snu})\right|\,,
\eeq
where $i{\cal M}_{nm}$ 
is the sum of all contributing one loop Feynman diagrams 
computed below and $m_{\snu}$ is the R-parity-conserving tree-level 
sneutrino mass. 
In the more complicated case, where there are $n_{\rm deg}$ flavors
of mass-degenerate sneutrinos,
sneutrino/antisneutrino mass-eigenstates
are obtained by diagonalizing the $2n_{\rm deg} \times 2n_{\rm deg}$ sneutrino 
squared-mass matrix:
\beq
M^2_{\rm sneutrino}=
\pmatrix{m^2_{\snu}\, \delta_{mn} & {\cal M}_{mp}(p^2=m^2_{\snu}) \cr 
{\cal M}^*_{qn}(p^2=m^2_{\snu}) & m^2_{\snu}\, \delta_{qp}} \,,
\eeq
where $m,n=1,\ldots, n_{\rm deg}$ and 
$p,q=n_{\rm deg}+1,\ldots,2n_{\rm deg}$.  
In the case that there are small mass-splittings between sneutrinos of
different flavor, we can treat such effects perturbatively by simply
including such flavor non-degeneracies in the diagonal blocks
above.  Likewise, a small tree-level splitting of the sneutrino and
antisneutrino can be accommodated perturbatively by an appropriate
modification of the off-diagonal blocks above.

As discussed in Section IV.A,
we need only consider in detail the contribution of lepton and slepton
loops.  (In particular, we neglect flavor mixing, but 
allow for mixing between the L-type and R-type charged sleptons 
separately in each generation.)
The corresponding contributions of the quark and squark loops
are then easily obtained by appropriate substitution of parameters.
The relevant graphs with an intermediate lepton and slepton loops are
shown in Figs.~2 and 3 respectively.  

\begin{center}
\begin{picture}(200,130)(0,0)
\ArrowArcn(100,60)(40,180,0)
\ArrowArcn(100,60)(40,0,180)
\DashArrowLine(10,60)(60,60)5
\DashArrowLine(190,60)(140,60)5
\Text(30,50)[]{$\snu_p(p)$}
\Text(160,50)[]{$\snu_q(p)$}
\Text(95,115)[]{$\ell^-_m(q)$}
\Text(95,9)[]{$\ell^-_n(q-p)$}
%
%
\end{picture}
\end{center}
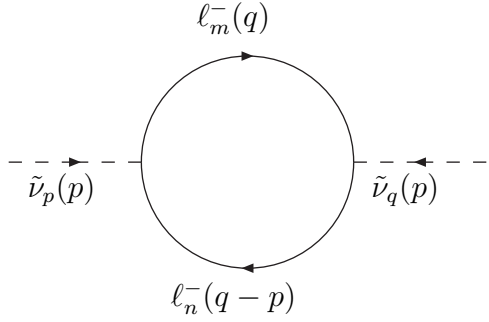
\centerline{Fig. 2. Lepton pair loop contribution to the
sneutrino--antisneutrino mass splitting.}
\vspace*{3mm}

%
\begin{center}
\begin{picture}(200,130)(0,0)
\DashArrowArcn(100,60)(40,180,0)5
\DashArrowArcn(100,60)(40,0,180)5
\DashArrowLine(10,60)(60,60)5
\DashArrowLine(190,60)(140,60)5
\Text(30,50)[]{$\snu_p(p)$}
\Text(160,50)[]{$\snu_q(p)$}
\Text(95,115)[]{$\widetilde \ell^-_m(q)$}
\Text(95,9)[]{$\widetilde \ell^-_n(q-p)$}
%
%
\end{picture}
\end{center}
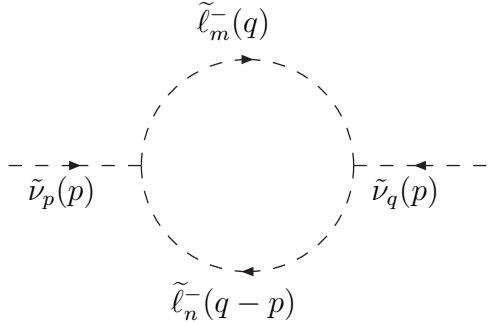
\centerline{Fig. 3. Slepton pair loop contribution to the
sneutrino--antisneutrino mass splitting.}
\vspace*{4mm}

Using the Feynman rules of Appendix C (including a minus sign for the
fermion loop), the contribution of the lepton loop (Fig.~2)
is given by
\beqa \label{mell}
i{\cal M}^{(f)}_{pq}&=&-\sum_{m,n}\lambda_{pmn} \lambda_{qnm} 
\int {d^4 q \over (2 \pi)^4}  
{{\rm Tr} \left[ (\qslash + m_m)P_L (\pslash+\qslash+m_n)P_L \right] 
\over [q^2-m_m^2][(q+p)^2-m_n^2]} \\
&=& 
{-i\over 8 \pi^2}\sum_{m,n} \lambda_{pmn} \lambda_{qnm} m_m m_n 
B_0(p^2,m_m^2,m_n^2).
\nonumber
\eeqa

The contribution of the slepton loop (Fig.~3) contains two distinct
pieces.  In the absence of LR slepton mixing, we have LL and
RR contributions in the loop proportional to the $\lambda$ Yukawa
couplings.  When we turn on the LR slepton mixing, 
we find additional contributions proportional to the corresponding
$A$-terms.  First, consider the contributions proportional to Yukawa
couplings.  For simplicity, we neglect the LR slepton mixing in
this case.  As before, we work in a basis where $v_m=0$ ({\it i.e.},
$v_0\equiv v_d$) and we choose a flavor basis
corresponding to the one where the charged lepton mass matrices are
diagonal.  Then, the contribution of the slepton loop (Fig.~3),
summing over $i=$L,R type sleptons is given by
\beqa \label{mlambda}
i{\cal M}^{(\lambda)}_{pq}&=&\sum_{i,m,n}\lambda_{pmn} \lambda_{qnm} m_m m_n 
\int {d^4 q \over (2 \pi)^4} 
{1 \over [q^2-M_{m_i}^2][(q+p)^2-M_{n_i}^2]} \\
&=& 
{i \over 16 \pi^2}\sum_{mn} \lambda_{pmn} \lambda_{qnm}  m_m m_n
\left[ B_0(p^2,M_{m_R}^2,M_{n_R}^2) +B_0(p^2,M_{m_L}^2,M_{n_L}^2) \right],
\nonumber
\eeqa
where the $m_n$ are {\it lepton} masses, and $M_{m_i}$ are slepton masses.
It is easy to check that the divergences cancel from the sum 
$i{\cal M}^{(f)}_{pq}+i{\cal M}^{(\lambda)}_{pq}$, which results in a
finite correction to the sneutrino mass.  This serves as an important
check of the calculation.

If LR slepton mixing is included, the above results are
modified.  
The corrections to \eq{mlambda} in this case are easily obtained,
but we shall omit their explicit form here.  In addition, new slepton loop
contributions arise that are proportional to the $A$-parameters
(defined in \eq{softsusy}).  We
quote only the final result:
\beqa \label{radsnumix}
i{\cal M}^{(A)}_{pq}
&=&{i\over 64 \pi^2}
\sum_{m,n} a_{pmn}a_{qnm} \sin 2\phi_m \sin 2\phi_n \\ 
&\times&\left[ B_0(p^2,M_{m_1}^2,M_{n_1}^2)+B_0(p^2,M_{m_2}^2,M_{n_2}^2)-
B_0(p^2,M_{m_1}^2,M_{n_2}^2)-B_0(p^2,M_{m_2}^2,M_{n_1}^2) \right], \nonumber
\eeqa
where $\phi_n$ is the slepton mixing angle of the n{\it th}
generation, and the corresponding slepton eigenstate masses are
$M_{n_1}$ and $M_{n_2}$.  This result is manifestly finite.
Note that this contribution vanishes when
the LR mixing is absent.  

The total contribution of the lepton and slepton loops
are given by the sum of \eqss{mell}{mlambda}{radsnumix}:
\beq
i{\cal M}^{(\ell)}_{pq}=i{\cal M}^{(f)}_{pq}+i{\cal M}^{(\lambda)}_{pq}+
i{\cal M}^{(A)}_{pq}\,.
\eeq
Finally, one must add the contributions of the quark and squark
loops.  The results of this subsection can be used, with the substitutions
described in Section IV.A to derive the final expressions.  Once
again, we see that for generic choices of the
$\lambda$, $A$, $\lambda'$ and $A'$ parameters, all 
sneutrino--antisneutrino pairs (including those 
pairs that were mass-degenerate at tree-level)
are split in mass by one loop effects.

\section{Phenomenological Consequences}

The detection of a non-vanishing sneutrino--antisneutrino mass splitting
would be a signal of lepton number violation.  In particular, it
serves as a probe of $\Delta L=2$ interactions, which also
contributes to the generation of neutrino masses.  Thus, sneutrino
phenomenology at colliders may provide access to physics that
previously could only be probed by observables sensitive to neutrino
masses.  

Some proposals for detecting the sneutrino--antisneutrino mass splitting
were presented in \Ref{GH}.
If this mass splitting is large (more then about 1$\,$GeV) one
may hope to be able to reconstruct the two masses in
sneutrino pair-production, and measure their difference. 
In an RPV theory with $L$-violation,
resonant production of sneutrinos become possible \cite{EFP}
and the sneutrino mass splitting may be detected either directly 
\cite{FGH} or by using tau-spin asymmetries \cite{Bar-Shalom}. 
If the mass splitting is much smaller than 1~GeV, 
sneutrino--antisneutrino oscillations can be used
to measure $\Delta m_{\snu}$.  In analogy with
$B$--$\overline B$  mixing, a same sign lepton 
signal will indicate that the two sneutrino mass eigenstates are not 
mass-degenerate.   In practice, one may only be able to measure 
the ratio $x_{\snu}\equiv \Delta m_{\snu}/\Gamma_{\snu}$.
In order to be able to observe the oscillation two conditions must by
satisfied: (i) $x_{\snu}$ should not be much smaller than 1; and 
(ii) the branching ratio into a lepton number 
tagging mode should be significant.

The sneutrino--antisneutrino mass splitting is proportional to the 
RPV parameters $b_m$ (for tree-level mass splitting) 
and $\lambda$, $A$, $\lambda'$ and $A'$
(for loop-induced mass splitting). Generally speaking, these parameters
can be rather large, and the strongest bounds on them come from the 
limits on neutrino masses. 
In the following discussion, we will consider the possible values of 
the relevant parameters: (i)~the ratio of the sneutrino--antisneutrino
mass splitting to the neutrino mass ($r_\nu\equiv\Delta\msnu/m_\nu$);
(ii)~the sneutrino width ($\Gamma_{\snu}$); and (iii)~the
branching ratio of the sneutrino into a lepton number tagging mode.

\subsection{Order of magnitude of $\Delta\msnu/m_\nu$}

To determine the order of magnitude of $\Delta\msnu/m_\nu$, we shall
take all R-parity-conserving supersymmetric parameters to be of order
$\mz$.  In the one generation model, the neutrino acquires a mass of order
$m_\nu\sim\mu_1^2\cos^2\beta/\mz$ via tree-level mixing, where we have
used $\sin\xi=\mu_1/\mu$ in a basis where $v_1=0$.  The tree-level
mass splitting of the sneutrino-antisneutrino pair is obtained from
\eq{dms}, and we find $\dmsnutwo\sim b_1^2\sin^2\beta/\mz^2$.  
Using $\dmsnutwo=2m_{\snu\snu^*}\Delta\msnu$, it follows that
\beq \label{rnuest}
r_\nu\equiv {\Delta\msnu\over m_\nu}\sim
{b_1^2\tan^2\beta\over\mz^2\mu_1^2}\,. 
\eeq 
To appreciate the implications of this result, we note that
\eq{veveq} in the $v_1=0$ basis yields 
\beq \label{bone}
b_1= [({M^2_{\tilde L}})_{10}+ \mu_1\mu_0]\cot\beta\,.
\eeq
The natural case is the one where all terms in \eq{bone} are of the
same order.  Then $b_1\sim {\cal O}(\mz\mu_1\cot\beta)$, and it
follows that $r_\nu\sim  {\cal O}(1)$.  On the other hand, 
it is possible to have $r_\nu\gg 1$ if, {\it e.g.},
$({M^2_{\tilde L}})_{10}\gg\mu_1\mu_0$.  The upper bound, $r_\nu\lsim
10^3$ [see \eq{rnulimits}] still applies in the absence of unnatural
cancellations between the tree-level and the one-loop contributions to
$m_\nu$.

We do not discuss here any models that predict the relative size
of the relevant RPV parameters. We only note that while 
we are not familiar with specific one-generation models that lead to
$r_\nu \gg 1$, we are aware of models that 
lead to $r_\nu \sim 1$.  One such example is a class of models based
on horizontal symmetry \cite{bgnn}.

In the three generation model, there is at most one tree-level non-zero
neutrino mass, while all sneutrino--antisneutrino pair masses may be
split.  This provides far greater freedom for the possible values of 
$(\Delta\msnu)_m\sim b_m^2\sin^2\beta/\mz^3$, 
since in many cases these are not constrained by the
very small neutrino masses.  In general, significant regions of
parameter space exist in which $r_\nu\gg 1$ for at least $n_g-1$
generations of neutrinos and sneutrinos.

Consider next the implications of the RPV one loop corrections.  
These are proportional to different RPV parameters as compared
to those that control the tree-level neutrino masses and
sneutrino--antisneutrino mass splittings.  Thus, one may envision cases
where the RPV one loop results are either negligible, of the same order, or
dominant with respect to the tree-level results.  If the RPV one loop
results are negligible, then the discussion above applies.  In
particular, in the three generation model with generic model
parameters, one typically expects
$r_\nu\sim{\cal O}(1)$ for one of the generations, while $r_\nu\gg 1$
for the other two generations.  
In contrast, if the RPV one loop corrections are dominant,
then the results of Section IV imply that $r_\nu\sim{\cal O}(1)$ for all
three generations, for generic model parameters.

\subsection{Sneutrino width and branching ratios}

Besides their effect on the sneutrino--antisneutrino mixing, 
the RPV interactions also
modify the sneutrino decays. This can happen in two ways. First, the
presence of the $\lambda$ and $\lambda'$ coupling can directly mediate
sneutrino decay to quark and/or lepton pairs. 
Second, the sneutrinos can decay through their mixing with the Higgs
bosons (which would favor the decay into the heaviest fermion or boson
pairs that are kinematically allowed).  These decays are relevant if
the sneutrino is the lightest supersymmetric particle (LSP), or if the
R-parity-conserving sneutrino decays are suppressed ({\it e.g.}, if no
two-body R-parity-conserving decays are kinematically allowed).  

Consider two limiting cases.  First, suppose that the RPV
decays of the sneutrino are dominant (or that the sneutrino is the LSP).
Then, in the absence of CP-violating effects, the sneutrino and
antisneutrino decay into the same channels with the same rate.
Moreover, the RPV sneutrino decays violate lepton number by one unit.
Hence, one cannot identify the decaying (anti)sneutrino state via a
lepton tag, as in \Ref{GH}.  However, oscillation phenomena may still      
be observable if there is a significant difference in the CP-even and
CP-odd sneutrino lifetimes.
For example, if the RPV sneutrino decays via Higgs mixing
dominate, then for sneutrino masses between $2m_W$ and $2m_t$, 
the dominant decay channels for the CP-even scalar would be $W^+W^-$,
$ZZ$ and $h^0 h^0$, while the CP-odd scalar would decay mainly
into $b\bar b$.   In this case, the ratio of sneutrino lifetimes would be of
order $m_Z^2/m_b^2$.  Adding up all channels, one finds a
ratio of lifetimes of order $10^3$.  Moreover, the overall lifetimes are 
suppressed by small RPV parameters, so one can imagine cases where an LSP
sneutrino would decay at colliders with a displaced vertex.
Oscillation phenomena similar to that of the $K$--$\overline K$
system would then be observable for the sneutrino--antisneutrino
system.  Including all three generations of sneutrinos would lead to a
very rich phenomenology that would provide a precision probe of the
underlying lepton-number violation of the theory.  

Second, suppose that the R-parity-conserving decays of the sneutrino
are dominant.  Then, the considerations of \Ref{GH} apply.  In
particular, in most cases, there are leptonic final states in
sneutrino decays that tag the initial sneutrino state.  Thus, the
like-sign dilepton signal of \Ref{GH} can be used to measure 
$x_{\snu}=\Delta m_{\snu}/\Gamma_{\snu}$.
Since only values of $x_{\snu}\gsim 1$ are practically measurable, 
the most favorable case corresponds to very small
$\Gamma_{\snu}$.  In typical models of R-parity-conserving
supersymmetry, the sneutrino decays into two body final states with a
width of order 1$\,$GeV.  This result can be suppressed somewhat by
chargino/neutralino mixing angle and phase space effects, but the
suppression factor is at most a factor of $10^4$ in rate (assuming
that the tagging mode is to be observable).  If the LSP is the
$\widetilde\tau^\pm$, then supersymmetric models can be envisioned
where two-body sneutrino decays are absent, and the three-body
sneutrino decays $\snu_\ell\to\tilde\tau_R\nu_\tau\ell$ can serve as
the tagging mode.  In \Ref{GH}, we noted that an LSP $\tilde\tau_R$ is  
strongly disfavored by astrophysical bounds on the abundance of stable
heavy charged particles \cite{staulsp}.  In R-parity-violating
supersymmetry, this is not an objection, since the LSP $\tilde\tau_R$
would decay through an RPV interaction.  Three-body sneutrino decay
widths can vary typically between 1$\,$eV and 1$\,$keV, depending on the
supersymmetric parameters.  Thus, in this case, the like-sign dilepton
signature can also provide a precision probe of the
underlying lepton-number violation of the theory. 

\subsection{Conclusions}

R-parity violating low-energy supersymmetry with baryon number conservation
provides a framework for
particle physics with lepton-number violation.  Recent experimental
signals of neutrino masses and mixing may provide the first glimpse of
the lepton-number violating world.  The search for neutrino masses and
oscillations is a difficult one.  Even if successful, such observations will
provide few hints as to the nature of the underlying lepton number
violation.  In supersymmetric models that incorporate lepton number
violation, the phenomenology of sneutrinos may provide additional
insight to help us unravel the mystery of neutrino masses and mixing.
Sneutrino flavor mixing and 
sneutrino--antisneutrino oscillations are analogous to neutrino 
flavor mixing and Majorana neutrino masses, respectively.
Crucial observables at future colliders include the
sneutrino--antisneutrino mass splitting, sneutrino oscillation
phenomena, and possible long sneutrino and antisneutrino lifetimes.
In this paper, we described CP-conserving sneutrino phenomenology
that can probe the physics of lepton number violation.  In a
subsequent paper, we will address the implications of CP-violation in
the sneutrino system.  The observation of such phenomena at future
colliders would have a dramatic impact on the pursuit of physics
beyond the Standard Model.

\acknowledgments
We thank Yossi Nir for helpful discussions.
YG is supported by the U.S. Department of Energy under
contract DE-AC03-76SF00515, and
HEH is supported in part by the U.S. Department of Energy
under contract DE-FG03-92ER40689 and in part by a Frontier Fellowship
from Fermi National Accelerator Laboratory.

\appendix

\section{The scalar potential}
In softly-broken supersymmetric theories, the total scalar potential 
is given by \eq{scalarpot}, where $V_F$ and $V_D$ originate from the
supersymmetry-preserving sector, while $V_{\rm soft}$ contains the
soft-supersymmetry-breaking terms.  $V_F$ is obtained from the
superpotential $W$ by first replacing all chiral superfields by their leading
scalar components and then computing
\beq \label{vsubf}
V_F = \sum_{\Phi} \left|{dW \over d \Phi}\right|^2,
\eeq
where the sum is taken over all contributing scalar fields, $\Phi$.
For the superpotential in \eq{rpvsuppot} we obtain:
\beqa
{dW \over d D_m} &=& \lp_{\alpha nm} L_\alpha^i Q_n^j \epsilon_{ij}, \\
{dW \over d U_m} &=& -h_{nm} H_U^i Q_n^j \epsilon_{ij}, \nonumber \\
{dW \over d Q_m^j} &=& \left(\lp_{\alpha nm} L_\alpha^i D_m - 
     h_{nm} H_U^i U_m \right)\epsilon_{ij}, \nonumber \\
{dW \over d E_m} &=& \half \l_{\alpha\beta m} 
     L_\alpha^i L_\beta^j \epsilon_{ij}, \nonumber \\
{dW \over d L_\alpha^i} &=& 
    \left(\l_{\alpha\beta m} L_\beta^j E_m + \lp_{\alpha nm} Q_n^j D_m 
    -\mu_\alpha H_U^j \right) \epsilon_{ij}, \nonumber \\
{dW \over d H_U} &=& \left(h_{nm} Q_n^i U_m
    -\mu_\alpha L_\alpha^i \right)\epsilon_{ij}. \nonumber 
\eeqa
Inserting these results into \eq{vsubf}, one ends up with:
\beqa \label{fpot}
V_F &=& 
  \lp_{\alpha nm} \lps_{\gamma km} L_\alpha^i Q_n^j 
  \left(L_\gamma^{i*} Q_k^{j*} - L_\gamma^{j*} Q_k^{i*}\right) 
   + h_{nm} h_{km}^* H_U^i Q_n^j 
 \left(H_U^{i*} Q_k^{j*} - H_U^{j*} Q_k^{i*} \right) \\
 &&\quad 
+\lp_{\alpha nm} \lps_{\gamma nk} L_\alpha^i L_\gamma^{i*}  D_m D_k^* + 
     h_{nm}h_{nk}^* |H_U|^2 U_m U_k^* \nonumber \\
&&\quad
  - (h_{nm} \lps_{\gamma nk} H_U^i L_\gamma^{i*} U_m D_k^* + {\rm h.c.})
 + \half \l_{\alpha\beta m} \ls_{\gamma\delta m} 
   L_\alpha^i L_\gamma^{i*} L_\beta^j L_\delta^{j*}
\nonumber \\
&&\quad
+ \l_{\alpha\beta m} \ls_{\alpha\gamma k} L_\beta^i L_\gamma^{i*} E_m E_k^* + 
\lp_{\alpha nm} \lps_{\alpha pk} Q_n^i Q_p^{i*} D_m D_k^* 
     \nonumber \\ &&\quad
+ |\mu_\alpha|^2 |H_U|^2 + 
(\l_{\alpha\beta m} \lps_{\alpha pk} L_\beta^i Q_p^{i*} E_m D_k^* +
   {\rm h.c.})
\nonumber \\ &&\quad
- (\mu_\alpha \ls_{\alpha\gamma k} H_U^i L_\gamma^{i*} E_k^* + {\rm h.c.})
- (\mu_\alpha \lps_{\alpha pk}  H_U^i Q_p^{i*} D_k^* + {\rm h.c.})
\nonumber \\
&&\quad
+ \mu_\alpha \mu_\beta^*  L_\alpha^i L_\beta^{i*} + 
h_{nm} h^*_{pq} U_m U^*_q Q_n^i Q_p^{i*} 
- (\mu_\alpha h^*_{pq} L_\alpha^i Q_p^{i*} U^*_q + {\rm h.c.})\,. \nonumber
\eeqa
$V_D$ is obtained from the following formula
\beq \label{vsubd}
V_D=\half \left[D^a D^a + (D')^2 \right]\,,
\eeq
where 
\beqa
D^a&=&\half g\left[
H_U^{i*} \sigma^a_{ij} H_U^j + \sum_m \widetilde Q_m^{i*}
\sigma^a_{ij} \widetilde Q_m^j  + \sum_\alpha \widetilde L_\alpha^{i*}
\sigma^a_{ij} \widetilde L_\alpha^j \right] \\
D'&=&\half g'\left[|H_U|^2 - \sum_\alpha |\widetilde L_\alpha|^2
+2 \sum_m |\widetilde E_m|^2 + {\textstyle{1 \over 3 }}
\sum_m |\widetilde Q_m|^2 - {\textstyle{4 \over 3 }}
\sum_m |\widetilde U_m|^2 + {\textstyle{2 \over 3 }}
\sum_m |\widetilde D_m|^2 \right]. \nonumber
\eeqa
Then, 
\beqa \label{dpot}
V_D= &&\eighth g^2 \Biggl\{
\left(|H_U|^2 -  \sum_\alpha |\widetilde L_\alpha|^2
- \sum_m |\widetilde Q_m|^2 \right)^2
-2 \sum_{\alpha\ne \beta} 
|\epsilon_{ij} \widetilde L_\alpha^i \widetilde L_\beta^j |^2
+4 \sum_\alpha |H_U^{i*} \widetilde L_\alpha^i|^2 \\ &&
-2 \sum_{m\ne n}  |\epsilon_{ij} \widetilde Q_m^i \widetilde Q_n^j |^2
+4 \sum_m |H_U^{i*} \widetilde Q_m^i|^2 
-4 \sum_{\alpha m} |\epsilon_{ij} \widetilde L_\alpha^i \widetilde Q_m^i|^2  
\Biggr\} \nonumber \\
&&+
\eighth g^{\prime 2}
\Biggl[|H_U|^2 - \sum_\alpha |\widetilde L_\alpha|^2
+2 \sum_m |\widetilde E_m|^2 + {\textstyle{1 \over 3 }}
\sum_m |\widetilde Q_m|^2 - {\textstyle{4 \over 3 }}
\sum_m |\widetilde U_m|^2 + {\textstyle{2 \over 3 }}
\sum_m |\widetilde D_m|^2 \Biggr]^2. \nonumber
\eeqa

Finally, the soft-supersymmetry-breaking contribution to the scalar
potential has already been given in \eq{softsusy}.

\section{The charged fermion and scalar sectors}

Using the same techniques discussed in Section III, one can evaluate
the tree-level masses of charged fermions and scalars.  For
completeness, we include here the results for the general R-parity-violating,
baryon-triality-preserving model exhibited in Section II.  (For
related results in a minimal RPV model in which $\mu_m$ is the only 
RPV parameter, see \Ref{minimalr}.)

First, we consider the sector of charged fermions.  The charginos and
charged leptons mix, so we must diagonalize a $(n_g+2)\times (n_g+2)$
matrix, for $n_g$ generations of leptons.  Following the notation
of \Ref{gh1}, we assemble the two-component fermion fields as follows:
\beqa
&&\psi^+=(-i\lambda^+,\psi^+_{H_U},\psi^+_{E_k})\,,\nonumber \\ 
&&\psi^-=(-i\lambda^-,\psi^-_{L_\alpha})\,,
\eeqa 
where $-i\lambda^\pm$ are the two component wino fields, and the
remaining fields are the fermionic components of the indicated scalar
field.   As before, $m=1,\ldots,n_g$ and $\alpha=0,1,\ldots,n_g$, with
$L_0\equiv H_D$.  The mass term in the Lagrangian then takes the 
form \cite{bgnn,enrico,akeroyd}:
\beq
{\cal L}_{\rm mass}=-\half(\psi^+\,\,\psi^-)\pmatrix{0&X^T\cr X&0\cr}
\pmatrix{\psi^+\cr \psi^-\cr}\,,
\eeq
where~\footnote{The result given in \eq{xmatrix} 
corrects a minor error that appears in \refs{bgnn}{enrico}.} 
\beq \label{xmatrix}
X=\pmatrix{M_2& \sqhalf gv_u & 0_m \cr
\sqhalf gv_\alpha & \mu_\alpha & (m_\ell)_{\alpha m}\cr}\,.
\eeq
In \eq{xmatrix}, $0_m$ is a row vector with $n_g$ zeros, and
\beq \label{mellgen}
(m_\ell)_{\alpha m}\equiv \sqhalf v_\rho\lambda_{\rho\alpha m}\,.
\eeq
Note that in the basis where $v_n=0$, the definition 
of $(m_\ell)_{nm}$ reduces to the one given in \eq{ellmassdef}.
The charged fermion masses
are obtained by either diagonalizing $X^\dagger X$ (with unitary matrix
$V$) or $X X^\dagger$ (with unitary matrix $U^*$), where the two 
unitary matrices are chosen such that $U^*X V^{-1}$ is a diagonal
matrix with the non-negative fermion masses along the diagonal.
The following relation is noteworthy:
\beq \label{tracerela}
{\rm Tr}~(X^\dagger X)=
{\rm Tr}~(XX^\dagger)=|M_2|^2+|\mu|^2+2m_W^2 +{\rm Tr}~(m^\dagger_\ell
m_\ell)\,,
\eeq
where $|\mu|^2$ is defined in \eq{mudef}.  Note that in the
R-parity-conserving MSSM,
${\rm Tr}~M^2_\chi\equiv |M_2|^2+|\mu|^2+2m_W^2$ is the sum
of the two chargino squared-masses and $m_\ell$ is the
charged lepton mass matrix.  In the presence of RPV
interactions, \eq{tracerela} remains valid despite the mixing 
between charginos and charged leptons.  Of course, $m_\ell$ no longer
corresponds precisely to a mass matrix of physical states.
For example, in the $v_m=0$ basis,
\beq \label{xdaggerx}
X^\dagger X=\pmatrix{|M_2|^2+\half g^2|v_d|^2 & \sqhalf 
g(M^*_2 v_u+v_d^*\mu\cos\xi) & 0_m \cr \sqhalf 
g(M_2 v_u^*+v_d\mu^*\cos\xi) & |\mu|^2+\half g^2|v_u|^2 &
\mu^*_n(m_\ell)_{nm}\cr 0_k & \mu_n(m^*_\ell)_{nk} & (m^\dagger_\ell
m_\ell)_{km}\cr}\,,
\eeq
where $\cos\xi$ is defined in \eq{xidef}.
As expected, if $\mu_m\neq 0$ (but small), then the physical
lepton eigenstates will have a small admixture of the charged 
higgsino eigenstate.  It is amusing to note that in the exact limit of
$m_\ell=0$, there are $n_g$ massless fermions
({\it i.e.}, the charged leptons), in spite of the mixing with the
charged higgsinos through the RPV terms.\footnote{It may seem from
\eq{xdaggerx} that the charged leptons are unmixed if $m_{\ell}=0$.  
But, one can
shown that this is not the case by computing $XX^\dagger$.  The mixing
originates from $\mu_m\neq 0$ appearing in the matrix $X$ [\eq{xmatrix}].} 


We next turn to the charged scalar sector.  In this case, the charged
sleptons mix with the charged Higgs boson and charged Goldstone boson
(which is absorbed by the $W^\pm$).  The resulting 
$(2n_g+2) \times (2n_g+2)$ squared mass-matrix can be obtained from the
scalar potential given by eqs.~(\ref{fpot}),
(\ref{dpot}) and (\ref{softsusy}).  
In the $\{H_U^1, \widetilde L_\beta^{2*}, \widetilde E_m\}$ basis,
the charged scalar squared-mass matrix is given by:
\beq
M_C^2=\pmatrix{m^2_{uu} +D & b^*_\beta+D_\beta & 
\mu^*_\beta (m_\ell)_{\beta m} \cr
b_\alpha+D^*_\alpha &  
m^2_{\alpha\beta} + (m_\ell m_\ell^\dagger)_{\alpha\beta} + D_{\alpha\beta} & 
\sqhalf (a_{\rho\alpha m} v_\rho - \mu^*_\rho \lambda_{\rho\alpha m}v^*_u)\cr
\mu_\alpha(m^*_\ell)_{\alpha k} & 
\sqhalf (a_{\rho\beta k}^* v_\rho^* - 
\mu_\rho\lambda_{\rho\beta k}^* v_u) & 
(M^2_{\widetilde E})_{km}+(m_\ell^\dagger m_\ell)_{km} + D_{km} \cr}\,, 
\eeq
where the matrix $m_\ell$ is defined in \eq{mellgen} and
\beqa \label{dfactors}
m^2_{uu}&\equiv & m^2_U+|\mu|^2\,,\\
m^2_{\alpha\beta}&\equiv & (M^2_{\widetilde L})_{\alpha\beta} 
+ \mu_\alpha \mu^*_\beta\,, \nonumber \\
D_{\alpha\beta}&\equiv & \quarter g^2 v^*_\alpha v_\beta
+\eighth(g^2-g^{\prime 2})(|v_u|^2-|v_d|^2)\delta_{\alpha\beta}\,,\nonumber \\
D_{km}&\equiv&\quarter g^{\prime 2}(|v_u|^2-|v_d|^2)\delta_{km}\,,\nonumber \\
D_\alpha&\equiv&\quarter g^2 v_\alpha v_u\,,\nonumber \\
D&\equiv&\eighth(g^2+g^{\prime 2})(|v_u|^2-|v_d|^2)+\quarter g^2 |v_d|^2\,.
\nonumber
\eeqa

As a check of the calculation, we have verified that $(-v_u,v_\beta^*,0)$
is an eigenvector of $M_C^2$ with zero eigenvalue, corresponding to
the charged Goldstone boson that is absorbed by the $W^\pm$.  The
computation makes use of the minimization conditions of the potential
[\eqs{mincondsa}{mincondsb}] and the antisymmetry of 
$\lambda_{\rho\beta k}$ and $a_{\rho\beta k}$
under the interchange of $\rho$ and $\beta$.

A useful sum rule can be derived in the CP-conserving limit.  We find:
\beq \label{chsumrule}
{\rm Tr}~M_C^2 = m_W^2+{\rm Tr}~M_{\rm odd}^2+{\rm Tr}~M_{\widetilde E}^2
+2\,{\rm Tr}~(m_\ell^\dagger m_\ell)
-\quarter n_g m_Z^2\cos2\beta\,.
\eeq
This is the generalization
of the well known sum rule, $m_{H^\pm}^2=m_W^2+m_A^2$, of the MSSM
Higgs sector \cite{hunter}.  
The charged sleptons are also contained in the above
sum rule.  As a check,
consider the one-generation R-parity-conserving MSSM limit.  Removing the
Higgs sum rule contribution from \eq{chsumrule}, the leftover pieces
are:
\beq \label{sleptonsum}
m^2_{\widetilde e_L}+m^2_{\widetilde e_R}-m^2_{\widetilde \nu}= 2m_e^2
+M_{\widetilde E}^2-\quarter m_Z^2\cos2\beta\,.
\eeq
The term in \eq{sleptonsum} that is proportional to $m_Z^2$
is simply the D-term contribution to the combination of
slepton squared-masses specified above.

\section{Feynman rules}

The fermion-scalar Yukawa couplings take the form:
\beq
{\cal L}_{\rm Yukawa}=-\half\left({\partial^2 W\over
\partial\phi_i\partial\phi_j}\right)\psi_i\psi_j+{\rm h.c.}\,,
\eeq
where superfields are replaced by their scalar components after taking
the second derivative of the superpotential $W$ [given in
\eq{rpvsuppot}], and the $\psi_i$ are two
component fermion fields.  Converting to four-component Feynman rules
(see, {\it e.g.}, the appendices of \Ref{haberkane}), and defining
$P_{R,L}\equiv\half(1\pm\gamma_5)$, we obtain the Feynman rules
listed in Fig.~4.  The charge conjugation matrix $C$ appears in
fermion-number-violating vertices.

\begin{center}
\begin{picture}(200,76)(0,0)
\DashArrowLine(10,40)(60,40)5
\ArrowLine(100,70)(60,40)
\ArrowLine(60,40)(100,10)
\Text(30,30)[]{$\snu_\alpha$}
\Text(70,20)[]{$e^-_\beta$}
\Text(70,67)[]{$e^-_m$}
\Text(140,40)[l]{$i\lambda_{\alpha\beta m}P_L$}
\end{picture}
\end{center}
\begin{center}
\begin{picture}(200,76)(0,0)
\DashArrowLine(60,40)(10,40)5
\ArrowLine(100,70)(60,40)
\ArrowLine(60,40)(100,10)
\Text(30,30)[]{$\snu_\alpha$}
\Text(70,20)[]{$e^-_\beta$}
\Text(70,67)[]{$e^-_m$}
\Text(140,40)[l]{$i\lambda^*_{\alpha\beta m}P_R$}
\end{picture}
\end{center}
\begin{center}
\begin{picture}(200,76)(0,0)
\DashArrowLine(10,40)(60,40)5
\ArrowLine(100,70)(60,40)
\ArrowLine(60,40)(100,10)
\Text(30,30)[]{$\widetilde e^-_{L_\alpha}$}
\Text(70,20)[]{$e^-_m$}
\Text(70,67)[]{$\nu_\beta$}
\Text(140,40)[l]{$i\lambda_{\alpha\beta m}P_L$}
\end{picture}
\end{center}
\begin{center}
\begin{picture}(200,76)(0,0)
\DashArrowLine(60,40)(10,40)5
\ArrowLine(100,70)(60,40)
\ArrowLine(100,10)(60,40)
\Text(30,30)[]{$\widetilde e^-_{R_m}$}
\Text(70,20)[]{$e^-_\beta$}
\Text(70,67)[]{$\nu_\alpha$}
\Text(140,40)[l]{$-i\lambda_{\alpha\beta m}C^{-1}P_L$}
\end{picture}
\end{center}
\begin{center}
\begin{picture}(200,76)(0,0)
\DashArrowLine(60,40)(10,40)5
\ArrowLine(60,40)(100,70)
\ArrowLine(100,10)(60,40)
\Text(30,30)[]{$\widetilde e^-_{L_\alpha}$}
\Text(70,20)[]{$e^-_m$}
\Text(70,67)[]{$\nu_\beta$}
\Text(140,40)[l]{$i\lambda^*_{\alpha\beta m}P_R$}
\end{picture}
\end{center}
\begin{center}
\begin{picture}(200,76)(0,0)
\DashArrowLine(10,40)(60,40)5
\ArrowLine(60,40)(100,70)
\ArrowLine(60,40)(100,10)
\Text(30,30)[]{$\widetilde e^-_{R_m}$}
\Text(70,20)[]{$e^-_\beta$}
\Text(70,67)[]{$\nu_\alpha$}
\Text(140,40)[l]{$i\lambda^*_{\alpha\beta m}P_R C$}
\end{picture}
\end{center}
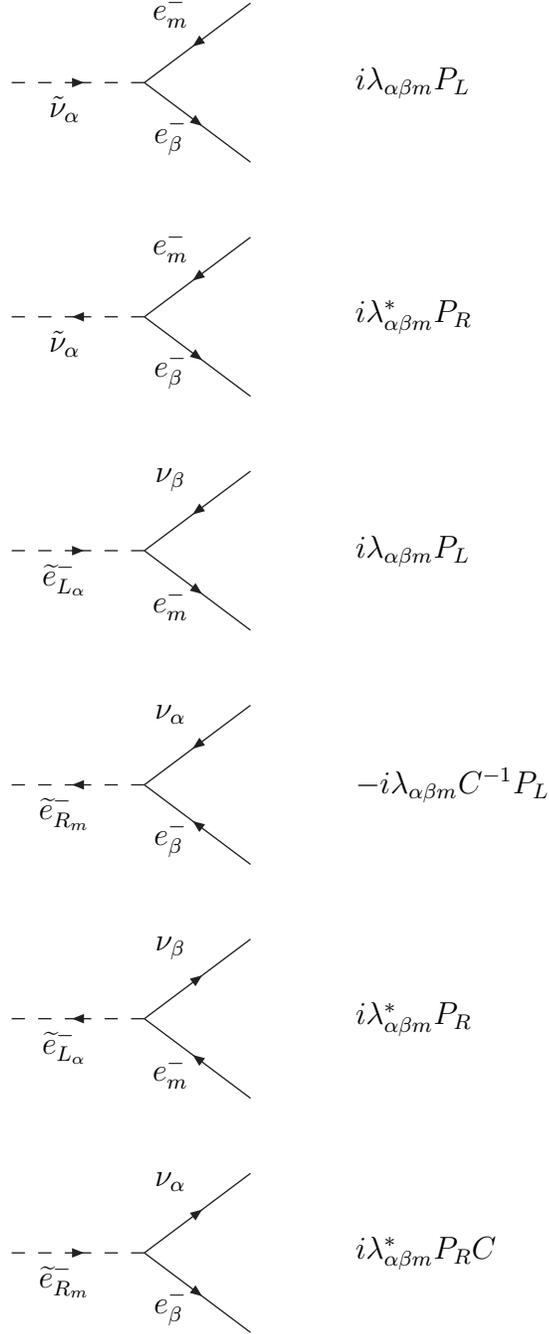
\centerline{Fig. 4. Feynman rules for the scalar--fermion interactions.}
\vspace*{3mm}

The Feynman rules for the cubic scalar interactions can be obtained from the
scalar potential [\eqss{fpot}{dpot}{softsusy}] by putting 
$\widetilde L^1_\alpha\to \widetilde L^1_\alpha +\sqhalf v_\alpha$.
The Feynman rules for the interaction of the sneutrinos 
with slepton pairs are given in Fig.~5, where $(m_\ell)_{\gamma m}$ is
defined in \eq{mellgen}.  In Section IV, we have applied the
rules of Fig.~5 to the ${\snu}_p \widetilde e_m \widetilde
e_n$ couplings ($p$, $m$, $n=1,\ldots,n_g$)
in the basis where $v_m=0$ and $(m_\ell)_{nm}$ is diagonal.
In this basis, the terms in Fig.~5 proportional to gauge couplings do
not contribute.

\begin{center}
\begin{picture}(200,76)(0,0)
\SetOffset(-100,0)
\DashArrowLine(10,40)(60,40)5
\DashArrowLine(100,70)(60,40)5
\DashArrowLine(60,40)(100,10)5
\Text(30,30)[]{$\snu_\alpha$}
\Text(70,20)[]{$\widetilde e^-_{R_n}$}
\Text(70,67)[]{$\widetilde e^-_{R_m}$}
\Text(140,40)[l]{$-i\lambda_{\alpha\gamma n}(m^*_\ell)_{\gamma m}
-{i\over 2\sqrt{2}}g^{\prime 2}v_\alpha^*\delta_{mn}$}
\end{picture}
\end{center}
\begin{center}
\begin{picture}(200,76)(0,0)
\SetOffset(-100,0)
\DashArrowLine(10,40)(60,40)5
\DashArrowLine(100,70)(60,40)5
\DashArrowLine(60,40)(100,10)5
\Text(30,30)[]{$\snu_\alpha$}
\Text(70,20)[]{$\widetilde e^-_{L_\rho}$}
\Text(70,67)[]{$\widetilde e^-_{L_\beta}$}
\Text(140,40)[l]{$-i\lambda_{\alpha\beta k}(m^*_\ell)_{\rho k}
+{i\over 4\sqrt{2}}\left[(g^2-g^{\prime 2})v_\alpha^*\delta_{\beta\rho}
-2g^2 v_\beta^*\delta_{\alpha\rho}\right]$}
\end{picture}
\end{center}
\begin{center}
\begin{picture}(200,76)(0,0)
\SetOffset(-100,0)
\DashArrowLine(10,40)(60,40)5
\DashArrowLine(100,70)(60,40)5
\DashArrowLine(60,40)(100,10)5
\Text(30,30)[]{$\snu_\alpha$}
\Text(70,20)[]{$\widetilde e^-_{R_n}$}
\Text(70,67)[]{$\widetilde e^-_{L_\beta}$}
\Text(140,40)[l]{$-ia_{\alpha\beta n}$}
\end{picture}
\end{center}
\vspace*{3mm}

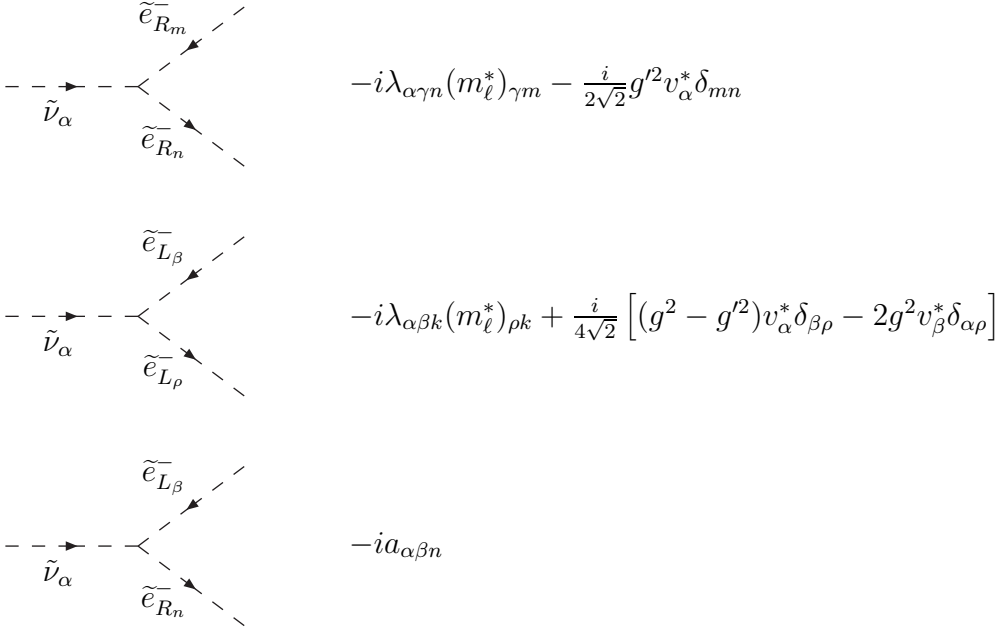
Fig. 5. Feynman rules for the interactions of the sneutrinos and 
charged sleptons.  

\section{The $B_0$ function}
The $B_0$ function is defined as follows:
\beq
{i \over 16 \pi^2} B_0(p^2,M^2,m^2) = 
\int {d^n q \over (2 \pi)^n} {1 \over (q^2-m^2)
\left[(q-p)^2-M^2\right]}\,.
\eeq
One can express $B_0$ as a one-dimensional integral:
\beq
B_0(p^2,M^2,m^2) = \Delta - \int_0^1 dx \ln \left(
{m^2 x + M^2 (1-x) - p^2 x(1-x) \over \mu^2}\right)\,,
\eeq
where 
\beq
\Delta \equiv (4 \pi)^\epsilon\, \Gamma(\epsilon)=
{1\over\epsilon}-\gamma+\ln(4\pi)+{\cal O}(\epsilon), \qquad 
\epsilon=2-{n \over 2}\,.
\eeq
Two limiting cases are useful for the calculations performed in
Section IV.  
In the $p^2 \to 0$ limit 
\beq
B_0(0,M_1^2,m^2) - B_0(0,M_2^2,m^2) = 
{M_2^2 \over m^2-M_2^2} \ln\left({m^2 \over M_2^2}\right) -
{M_1^2 \over m^2-M_1^2} \ln\left({m^2 \over M_1^2}\right)\,.
\eeq
If we furthermore take the $m \to 0$ limit, we obtain: 
\beq
B_0(0,M_1^2,0) - B_0(0,M_2^2,0) = \ln\left({M_1^2 \over M_2^2}\right)\,.
\eeq

{\tighten

}

\end{document}